# 1D-CNN-IDS: 1D CNN-based Intrusion Detection System for IIoT

Muhammad Arslan
Department of Computer Science
Loyola University
Chicago Illinois, United States
Thearslan380@gmail.com

Muhammad Mubeen
Department of Computer Science
University of People
Pasadena United States
Themubeenahmad334@gmail.com

Muhammad Bilal
Department of Artificial Intelligence
Rare Sense Inc
Covina, California, United States
Bilal@raresense.so

Saadullah Farooq Abbasi
Department of Electronic, Electrical and Systems Engineering
University of Birmingham
Birmingham, United Kingdom
S.f.abbasi@bham.ac.uk

*Abstract*— The demand of the Internet of Things (IoT) has witnessed exponential growth. These progresses are made possible by the technological advancements in artificial intelligence, cloud computing, and edge computing. However, these advancements exhibit multiple challenges, including cyber threats, security and privacy concerns, and the risk of potential financial losses. For this reason, this study developed a computationally inexpensive one-dimensional convolutional neural network (1DCNN) algorithm for cyber-attack classification. The proposed study achieved an accuracy of 99.90% to classify nine cyber-attacks. Multiple other performance metrices have been evaluated to validate the efficacy of the proposed scheme. In addition, comparison has been done with existing state-of-the-art schemes. The findings of the proposed study can significantly contribute to the development of secure intrusion detection for IIoT systems.

## I. INTRODUCTION

In recent years, there has been a remarkable increase in the use of the Internet of Things (IoT) [1]. This transformation has significantly raised the intelligence of manufacturing operations by integrating cutting-edge technologies, including artificial intelligence (AI), cloud computing, edge computing, big data analytics, robotics, and cybersecurity. The Industrial Internet of Things (IIoT) has notably augmented the level of automation within smart factories and distribution centers [2]. The IIoT serves as an extensive network connecting devices and providing computing services to smart industries, enhancing operational efficiency across sectors like manufacturing and service supply. However, its diverse nature exposes IIoT devices and networks to various cyber threats, posing significant security and privacy risks and potential financial losses [3]. Therefore, robust cybersecurity measures are essential to fully harness the benefits of IIoT.

Smart industries face a range of cyber threats, including unauthorized data access, IoT service disruption, and equipment damage [4]. Intrusion detection systems (IDSs) are crucial secondary defense mechanisms, working alongside other security measures to mitigate these risks. Utilizing traditional machine learning (ML) and deep learning (DL) techniques, modern IDSs analyze IoT traffic, detect attack patterns, and identify malicious activities. Yet, the dynamic nature of IIoT systems necessitates adaptive approaches beyond predefined attack classifications, while addressing data imbalance in cybersecurity datasets is vital for enhancing IDS performance and effectiveness.

CNN has been extensively used in the field of image classification [5], time series data classification [6] and text classification [7]. This study proposed a 1D CNN-based intrusion detection algorithm that is computationally inexpensive and takes less time for classification. The proposed algorithm used 3 CNN layers in addition to max pooling layers and dropout. After parameters optimization, the proposed study achieved an accuracy of 99.90%. The main contributions of the proposed study are as follows:

• The Edge-IIoT dataset has been pre-processed, which involves eliminating redundant feature columns and encoding categorical data. Subsequently, the fully pre-processed dataset is converted into a csv file, facilitating identification of different cyberattack patterns.

• An efficient 1DCNN algorithm has been developed to detect number of cyberattacks in IIoT.

• Rigorous experimentation has been done to evaluate the efficiency of the proposed 1D-CNN algorithm.

Remainder of the paper is organized in the following manner: Section II presents the brief literature of the existing studies in IIoT. Section III proposed the methodology whereas section IV presents the results. Finally, Section V concludes the paper.

## II. RELATED WORK

Over the past decade, traditional intrusion detection systems (IDSs) have faced evolving challenges in terms of their precision, efficacy, and resilience. In response to these challenges, Li et al. [8] proposed an algorithm for intrusion detection within Internet of Vehicles (IoV) environments. This approach incorporated a local update mechanism to generate labels from unknown data in emerging attacks. Experimental findings showed that the proposed study achieved a notable accuracy of 92%.

In an effort to further elevate classification accuracy, Mehedi et al. [9] Propounded a domain transfer learning (DTL)-based study tailored for IoT applications. This study emphasized efficient attribute selection and robust evaluation of a DTL-based ResNet model using real-world data. Gou et al. [10] proposed a distributed TL approach for IDSs, integrating TL into boosting algorithms to facilitate attack learning particularly in instances of suboptimal performance.

Their proposed framework demonstrated an impressive attack detection accuracy of 97.3%.

Furthermore, Mehedi et al. [11] engineered an IDS designed for In-Vehicle systems, incorporating a novel methodology to identify threat messages and accurately discriminate between normal and abnormal activities leveraging a LeNet model. Evaluation on a proprietary synthetic dataset yielded a commendable accuracy of 98.10%. The evolution of DTL and its consequential impacts across diverse domains have catalyzed the advancement of robust solutions for IIoT. Bierbrauer et al. [12] introduced an IDS using raw architecture traffic. Their approach trained a one-dimensional CNN in conjunction with a random forest model. The proposed study improved the overall classification accuracy upto 96.89%.

Xu et al. [13] propounded a secure intrusion detection system, employing an encryption scheme to secure classification models. Subsequently, XGBoost-based algorithm showcased effective model transfer with classification accuracy of 93.01%. Singh et al. [14] introduced an algorithm to detect darknet networks, employing a transformation of 1D features into 3D for enhanced accuracy. Abosata et al. [15] proposed a TL model for IoT IDSs, encompassing three primary stages. The proposed model exhibited high RPL security and an impressive accuracy of 85.52%.

Majority of the work presented recently has been trained and tested using 2-dimensional data. The researchers converted the data from 1D to 2D and then applied classification algorithms. However, this technique is computationally inefficient and expensive. For this reason, this study proposed a 1D CNN based algorithm that is faster than the existing algorithms and is computationally inexpensive.

III. METHODOLOGY

This section briefly explains the dataset, preprocessing, the proposed 1D-CNN, the evaluation parameters, and validation technique used in this study.

*A. Dataset*

The proposed study used Edge-IIoTset cybersecurity dataset that is publicly available from Ferrag et al. for further research [16]. The proposed dataset has been developed by generating a realistic IIoT environment with multiple IoT sensors including soil moisture, pH, heart rate, temperature, humidity and ultrasonic sensors. For the proposed study, we have used 10 classes to analyze the proposed algorithm. Fig 1. illustrates the class distribution of the proposed dataset.

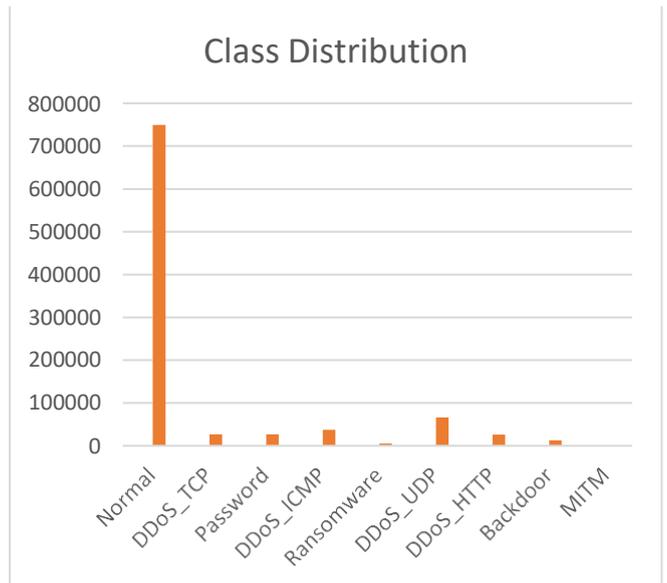

*Figure 1. Class distribution of the dataset used in proposed study.*

*B. Preprocessing*

Before training and testing the architecture, one of the most crucial step is to clean the dataset. The proposed dataset contains 62-features and 9 classes. The first step was to randomly select 10 classes. Then, NaN values have been removed from the dataset. Secondly, duplicate rows have been removed from the data. The new dataset contains 953239 rows with 62 columns.

*C. Proposed 1D CNN*

CNN is widely used for image recognition and classification in 100s of applications. Figure 2. shows the proposed 1D CNN architecture. The proposed 1DCNN architecture is designed to effectively extract features from time-series data. The methodology involves a detailed description of the architectural layers and components. This includes three convolutional layers that use 1D convolutions to extract local patterns and features from the input time-series data.

Following each convolutional layer, max-pooling layers are employed to downsample the feature maps, reducing their spatial dimensions while retaining the most salient features. Max-pooling helps in capturing invariant features and reducing computational complexity by discarding redundant information. The output from the final max-pooling layer is flattened and passed through two fully connected layers for further feature transformation and classification. These layers learn high-level representations of the input data and perform the classification task.

To prevent overfitting and improve generalization performance, a dropout layer has been used between the fully connected layers. Dropout randomly masks a fraction of the neurons, forcing the network to learn robust features that are not dependent on specific neurons. The methodology emphasizes the hierarchical nature of feature extraction in the 1DCNN architecture, where lower layers capture low-level temporal patterns, and higher layers learn more abstract representations. This hierarchical feature extraction process enables the network to effectively model complex temporal dynamics present in time-series data.

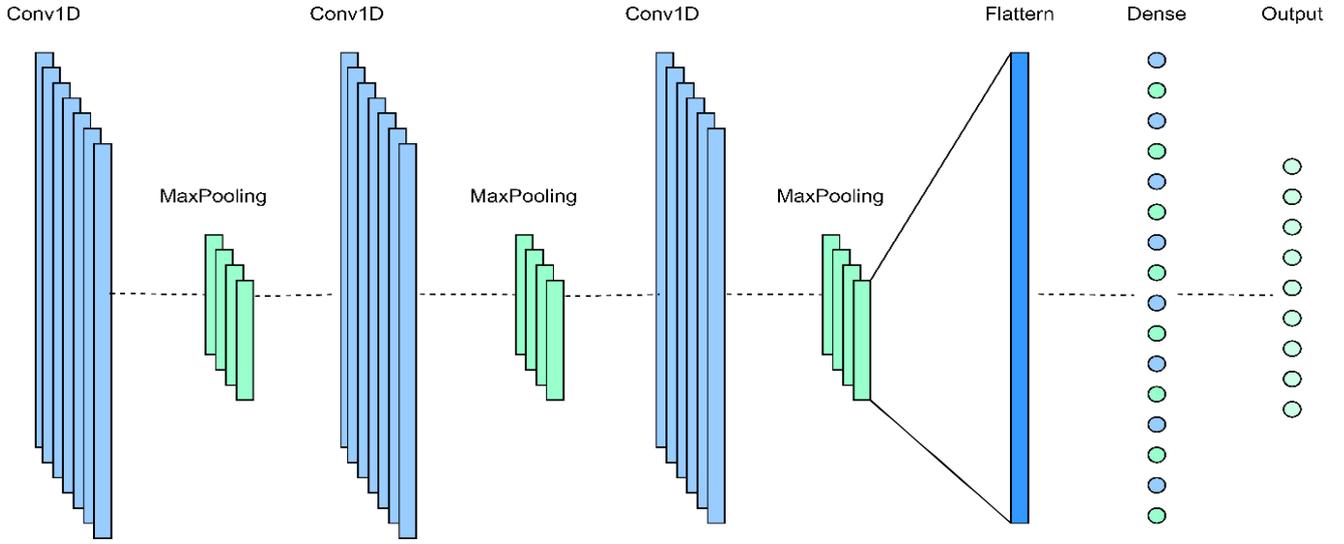

*Figure 2. The proposed 1D-CNN network architecture.*

Experimental validation of the proposed methodology involves training and testing the 1DCNN architecture with multiple performance metrices. Performance metrics like accuracy, precision, recall, and F1-score have been evaluated to evaluate the effectiveness of the proposed methodology. Additionally, comparisons with existing architectures have been conducted to demonstrate the superiority of the 1DCNN architecture in capturing complex temporal patterns from time-series data.

## IV. RESULTS

The proposed study has been trained and testing on Python 3.7 in 12th Gen Intel Core i9-12900H 2.50 GHz having Nvidia GPU. The proposed study achieved a mean accuracy of 99.90% for 9 classes. Figure shows the confusion matrix of the proposed study. In addition, tab shows the performance metrices of the proposed study calculated using the confusion matrix shown in fig 3.

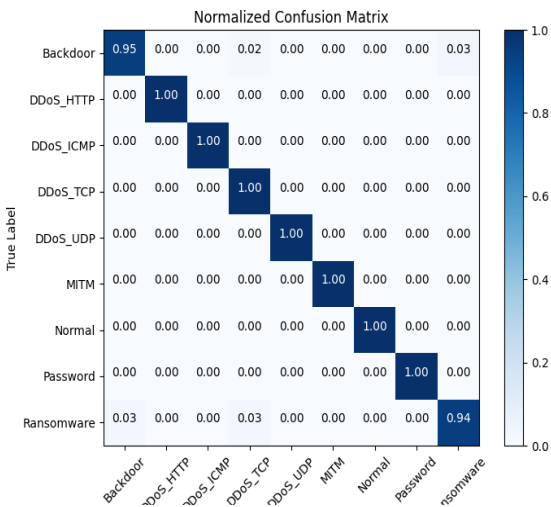

*Figure 3. Confusion matrix of the proposed study.*

*Table 1. Performance metrices of the proposed study.*

|  | Accuracy | Precision | Recall | F1-score |
|---|---|---|---|---|
| This Study | 99.90 | 98.80 | 98.79 | 98.78 |

In addition, figure 4 shows the training and validation loss with respect to the number of epochs. From the figure, it is evident that both training and validation loss minimizes at epoch 3 therefore we have used 3 epochs for training. Another important parameter to assess the efficiency of the proposed 1D CNN network is time. For training and validation, the proposed study requires 108.46 seconds whereas only 3.94 seconds are required to test the testing dataset. This proves that proposed study is time efficient and can process huge datasets in less than 4 seconds.

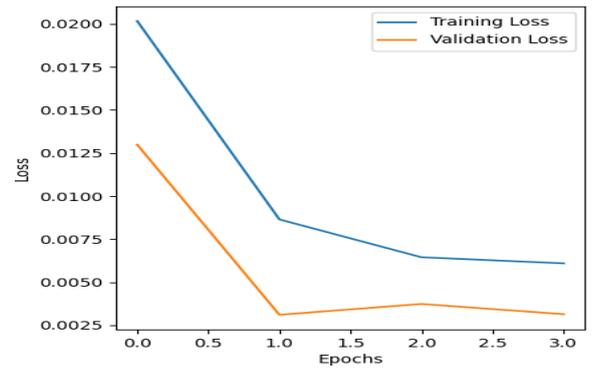

*Figure 4. Epochs vs loss graph.*

To evaluate the efficacy of the proposed architecture, it has been compared with latest intrusion detection schemes. Most of the algorithms are trained and tested on limited classes however the proposed study used 9 attacks for classification. Figure 5. shows the comparison of the proposed study with existing literature. Here, it is very important to note that the proposed study can classify 9 attacks on the IIoT environment however all the other algorithms classify less that 9 attacks. Therefore, this proves that the proposed study can classify more attacks in addition to the overall accuracy achieved.

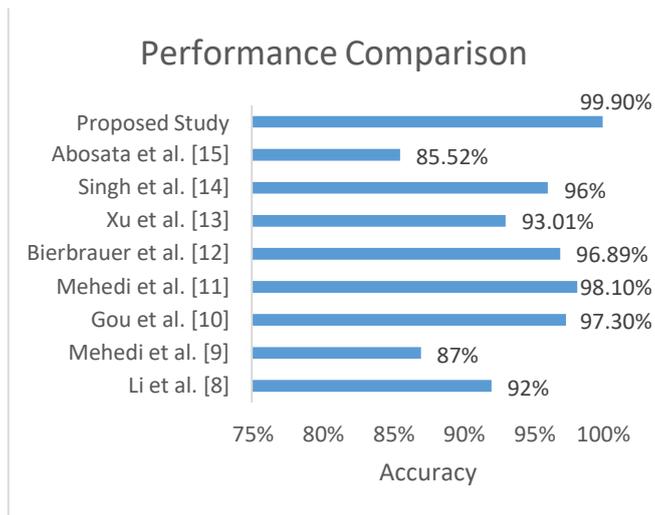

*Figure 5. Performance comparison of the proposed study.*

The proposed algorithm achieved remarkable results for intrusion detection in a real-world environment however there are still some limitations that needs to be addressed in the future studies. Firstly, there are 8 attacks that have been classified in the proposed study however there are multiple other attacks, i.e., Vulnerability Scanner, SQL injection, Port Scanning etc., that needs to be classified for broader applications. Additionally, in the proposed study, there is a class imbalance problem as there are some classes with huge data however some classes have limited data. This class imbalance problem needs to be addressed in future studies either by data augmentation or synthetic data generation techniques.

## V. Conclusion

This study proposed a new 1DCNN algorithm for cyberattack classification on IIoT data. Initially, NaN and duplicate values were removed to preprocess the data. Then, 1DCNN has been developed to test and train the Edge-IIoTset. The proposed architecture achieved an accuracy of 99.90% for the classification of nine attacks. Rigorous experimentation and validation have been used to evaluate the efficiency of the proposed scheme. The proposed study outperformed the existing studies for cyberthreat detection. In addition, 1DCNN is computationally inexpensive and does not require transforming the time series data into images. To conclude, it is evident that the proposed study can be used to detect real-time cyberthreats in industry 4.0.

## Acknowledgment

The authors would like to thank University of Birmingham, United Kingdom for providing the facilities to conduct this research.

## References

[1] Lampropoulos, G., Siakas, K. and Anastasiadis, T., 2019. Internet of things in the context of industry 4.0: An overview. International Journal of Entrepreneurial Knowledge, pp.4-19.

[2] Munirathinam, S., 2020. Industry 4.0: Industrial internet of things (IIOT). In Advances in computers (Vol. 117, No. 1, pp. 129-164). Elsevier.

[3] Sengupta, J., Ruj, S. and Bit, S.D., 2020. A comprehensive survey on attacks, security issues and blockchain solutions for IoT and IIoT. Journal of network and computer applications, 149, p.102481.

[4] Bécue, A., Praça, I. and Gama, J., 2021. Artificial intelligence, cyber-threats and Industry 4.0: Challenges and opportunities. Artificial Intelligence Review, 54(5), pp.3849-3886.

[5] Awais, M., Long, X., Yin, B., Abbasi, S.F., Akbarzadeh, S., Lu, C., Wang, X., Wang, L., Zhang, J., Dudink, J. and Chen, W., 2021. A hybrid DCNN-SVM model for classifying neonatal sleep and wake states based on facial expressions in video. IEEE Journal of Biomedical and Health Informatics, 25(5), pp.1441-1449.

[6] Abbasi, S.F., Abbasi, Q.H., Saeed, F. and Alghamdi, N.S., 2023. A convolutional neural network-based decision support system for neonatal quiet sleep detection. Mathematical Biosciences and Engineering, 20(9), pp.17018-17036.

[7] Tan, Z., Chen, J., Kang, Q., Zhou, M., Abusorrah, A. and Sedraoui, K., 2021. Dynamic embedding projection-gated convolutional neural networks for text classification. IEEE Transactions on Neural Networks and Learning Systems, 33(3), pp.973-982.

[8] Li, X., Hu, Z., Xu, M., Wang, Y. and Ma, J., 2021. Transfer learning based intrusion detection scheme for Internet of vehicles. Information Sciences, 547, pp.119-135.

[9] Mehedi, S.T., Anwar, A., Rahman, Z., Ahmed, K. and Islam, R., 2022. Dependable intrusion detection system for IoT: A deep transfer learning based approach. IEEE Transactions on Industrial Informatics, 19(1), pp.1006-1017.

[10] Gou, S., Wang, Y., Jiao, L., Feng, J. and Yao, Y., 2009, August. Distributed transfer network learning based intrusion detection. In 2009 IEEE International Symposium on Parallel and Distributed Processing with Applications (pp. 511-515). IEEE.

[11] Mehedi, S.T., Anwar, A., Rahman, Z. and Ahmed, K., 2021. Deep transfer learning based intrusion detection system for electric vehicular networks. Sensors, 21(14), p.4736.

[12] Bierbrauer, D.A., De Lucia, M.J., Reddy, K., Maxwell, P. and Bastian, N.D., 2023. Transfer learning for raw network traffic detection. Expert Systems with Applications, 211, p.118641.

[13] Xu, M., Li, X., Wang, Y., Luo, B. and Guo, J., 2021. Privacy - preserving multisource transfer learning in intrusion detection system. Transactions on Emerging Telecommunications Technologies, 32(5), p.e3957.

[14] Singh, D., Shukla, A. and Sajwan, M., 2021. Deep transfer learning framework for the identification of malicious activities to combat cyberattack. Future Generation Computer Systems, 125, pp.687-697.

[15] Abosata, N., Al-Rubaye, S. and Inalhan, G., 2022. Customised intrusion detection for an industrial IoT heterogeneous network based on machine learning algorithms called FTL-CID. Sensors, 23(1), p.321.

[16] Ferrag, M.A., Friha, O., Hamouda, D., Maglaras, L. and Janicke, H., 2022. Edge-IIoTset: A new comprehensive realistic cyber security dataset of IoT and IIoT applications for centralized and federated learning. IEEE Access, 10, pp.40281-40306.

[17] Arslan, M., Mubeen, M., Akram, A., Abbasi, S.F., Ali, M.S., and Tariq, M.U., 2024. A Deep Features-Based Approach Using Modified ResNet50 and Gradient Boosting for Visual Sentiments Classification. arXiv preprint arXiv:2408.07922. Available at: https://arxiv.org/abs/2408.07922.